\documentclass[12pt]{article}
\usepackage{amsfonts}
\usepackage{amsmath,amssymb}
\usepackage[nosort]{cite}
\usepackage[margin=1in]{geometry}
\usepackage{graphicx}

\makeatletter \@addtoreset{equation}{section} \makeatother

\makeatletter
\newcommand\blfootnote{\xdef\@thefnmark{}\@footnotetext}
\makeatother

\newcommand{\beq}{\begin{equation}}
\newcommand{\eeq}{\end{equation}}
\newcommand{\bea}{\begin{eqnarray}}
\newcommand{\eea}{\end{eqnarray}}

\input{tcilatex}

\begin{document}


\begin{titlepage}

%

\vspace*{.2cm}

\begin{center}

\vspace*{1cm}

{\Large\bf Relation between large dimension operators
\\ \vspace{0.3cm}
and oscillator algebra of Young diagrams  }


\vspace*{.5cm%
}

\vspace*{1.5cm}

 {\bf Hai Lin }


\vspace*{1cm}
\small{

{\it Department of Physics, Harvard University, MA 02138, USA \\
Department of Mathematics, Harvard University, MA 02138, USA \\ }

}

\end{center}

\vspace*{.2cm}%

\vspace*{2cm} \large
\begin{abstract}

\vspace*{0.5cm}

{\large }

The operators with large scaling dimensions can be labelled by
Young diagrams. Among other bases, the operators using restricted
Schur polynomials have been known to have a large $N$ but
nonplanar limit under which they map to states of a system of
harmonic oscillators. We analyze the oscillator algebra acting on
pairs of long rows or long columns in the Young diagrams of the
operators. The oscillator algebra can be reached by a Inonu-Wigner
contraction of the $u(2)$ algebra inside of the $u(p)$ algebra of
$p$ giant gravitons. We present evidences that integrability in
this case can persist at higher loops due to the presence of the
oscillator algebra which is expected to be robust under loop
corrections in the nonplanar large $N$ limit.

\end{abstract}

\end{titlepage}


\setcounter{tocdepth}{2}


\section{Introduction}

\label{sec: introduction}

The AdS/CFT duality provides insights for both gauge theory and gravity
theory \cite{Maldacena:1997re,Gubser:1998bc,Witten:1998qj}. There are
interesting maps between the gauge side and the gravity side. For example,
the BMN operators \cite{Berenstein:2002jq} and spin chain operators \cite%
{Minahan:2002ve,Beisert:2003tq} are dual to string states on the gravity
side.

The large dimension operators can describe giant gravitons which
are branes on the gravity side
\cite{McGreevy:2000cw,Grisaru:2000zn,Hashimoto:2000zp,
Corley:2001zk,Balasubramanian:2001nh,Berenstein:2004kk}. The Schur
polynomial operators \cite{Corley:2001zk} labelled by Young
diagrams are dual to the giant graviton states on the gravity side
\cite{Corley:2001zk}. For instance, long rows of the Young
diagrams correspond to giant gravitons that grow in external
spacetime directions. Similarly, long columns of the Young
diagrams correspond to giant gravitons that grow in internal
directions.

It is interesting that the operators with large scaling dimensions can take
the form in the bases labelled by Young diagrams \cite%
{Corley:2001zk,Brown:2007xh,Kimura:2007wy,Bhattacharyya:2008rb,
Brown:2008ij,Kimura:2010tx}. These large dimension operators can be mapped
to both giant gravitons and bubbling geometries on the gravity side. For
example, a family of BPS operators can be labelled by the representations of
Brauer algebras \cite{Kimura:2007wy, Kimura:2010tx} and are connected to the
bubbling geometries \cite{Kimura:2011df}. The restricted Schur polynomial
operators and the operators with global symmetry basis or flavor basis can
conveniently describe the giant graviton states \cite%
{Brown:2007xh,Bhattacharyya:2008rb,Brown:2008ij}. The Brauer basis, the
restricted Schur basis, and the global symmetry basis or flavor basis, can
be transformed between each other.

The spectra of the operators describing giant graviton excitations have been
recently computed, for example \cite%
{Koch:2011hb,DeComarmond:2010ie,Carlson:2011hy,deMelloKoch:2011ci,
deMelloKoch:2012ck,deMelloKoch:2012sv,deMelloKoch:2011vn,
Koch:2010gp,Koch:2012sf,Koch:2011jk,Koch:2013xaa}. In a large $N$ but
nonplanar limit, integrability in the nonplanar regime were observed. There
have been many evidences of it at one and two loops in various sectors.

The operators can be diagonalized by harmonic oscillator states, for example
\cite{Koch:2011hb,Carlson:2011hy,deMelloKoch:2011ci}, and by double coset
ansatz \cite{deMelloKoch:2012ck}. The set of operators map to a system of
harmonic oscillators. The harmonic oscillator dynamics can be interpreted as
resulting from strings stretching between pairs of giant gravitons.

The picture of strings stretching between pairs of giant gravitons is
reminiscent of the approach in the eigenvalue picture of \cite%
{Berenstein:2005aa}, where the dynamics of background geometries and their
string excitations can be treated in an eigenvalue basis.

In this paper we analyze an oscillator algebra and its role in integrability
in the large $N$ but nonplanar regime. We study the relation between the
oscillator algebra and higher loop dilatation operators. We find that in the
nonplanar large $N$ limit the higher loop dilatation operators will not
correct the diagonalization of the operators if they satisfy the oscillator
algebra.

The organization of this paper is as follows. After introducing the relevant
bases of the operators and their mixing in section \ref{sec: operator}, we
analyze in section \ref{sec: su(2) contraction} the oscillator algebra
associated to the spectra of the operators, and their relation by
Inonu-Wigner contraction to an $u(2)$ algebra inside of an $u(p)$ algebra of
the Young diagram operators with $p$ long rows or long columns. In section %
\ref{sec: oscillator algebra}, we discuss the influence of the oscillator
algebra on the action of the higher loop dilatation operators. We find that
the $h$-loop dilatation operators preserve the integrability in the large $N$
but nonplanar limit, if they satisfy the oscillator algebra. We discuss
general number of pairs of giant gravitons or long rows in section \ref{sec:
general pairs}, and discuss the effective spring constants between them and
the influence of the higher loop corrections on the spectra. Finally we
briefly conclude in section \ref{sec: discussion}.

\vspace{1pt}

\vspace{1pt}

\vspace{1pt}

\vspace{1pt}

\vspace{1pt}


\section{Bases of operators and mixing of operators}

\vspace{1pt}\label{sec: operator}

\vspace{1pt}\vspace{0.1cm}\vspace{1pt}

\vspace{1pt} We will analyze restricted Schur polynomials built from the
fields in $\mathcal{N}$ = 4 gauge theory. It contains various interesting
sectors, such as the $su(2)$ sector and the $su(2|3)$ sector. It has several
scalar fields, and the Hermitian matrix scalars can be organized into
complex fields, for example
\begin{equation*}
Z=\phi ^{1}+i\phi ^{2},~~Y=\phi ^{3}+i\phi ^{4},\ ~X=\phi ^{5}+i\phi ^{6}.
\end{equation*}%
We will focus on restricted Schur polynomials built using $n$ $Z$ and $m$ $Y$
fields and will often refer to the $Y$ fields as \textquotedblleft
impurities\textquotedblright . The size $n+m$~is of order $O(N).$

\vspace{1pt}\vspace{0.1cm}These operators that we study have a large scaling
dimension. The restricted Schur polynomial in this case is, see for example
\cite{Bhattacharyya:2008rb},
\begin{equation}
\chi _{R,(r,s)\alpha \beta }(Z^{\otimes \,n},Y^{\otimes \,m})={\frac{1}{n!m!}%
}\sum_{\sigma \in S_{n+m}}\mathrm{Tr}(P_{R\rightarrow (r,s)\alpha \beta
}\Gamma _{R}(\sigma ))Z_{i_{\sigma (1)}}^{i_{1}}\cdots Z_{i_{\sigma
(n)}}^{i_{n}}Y_{i_{\sigma (n+1)}}^{i_{n+1}}\cdots Y_{i_{\sigma
(n+m)}}^{i_{n+m}}\,.
\end{equation}%
The label $R$ is an irreducible representation of the symmetric group $%
S_{n+m}$ in the form of a Young diagram with $n+m$ boxes. The labels $r$ and
$s$ are Young diagrams with $n$ and $m$ boxes respectively. The $r$ is an
irreducible representation of the group $S_{n}$ and the $s$ is an
irreducible representation of $S_{m}$. The group $S_{n+m}~$has a subgroup $%
S_{n}\times S_{m}$ whose irreducible representations are labelled by $(r,s)$%
. An irreducible representation $R~$of $S_{n+m}$ can subduce many different
representations $(r,s)~$of $S_{n}\times S_{m}$. The $\alpha \beta $ are
multiplicity labels of the irreducible representations $(r,s)$, which label
different ways that $(r,s)$ are subduced from $R$. The trace is realized by
including a projector $P_{R\rightarrow (r,s)}=P_{R\rightarrow (r,s)\alpha
\beta }$ and tracing over all of $R$, that is, $\mathrm{Tr}(P_{R\rightarrow
(r,s)\alpha \beta }\Gamma _{R}(\sigma ))$. This projector is from the
carrier space of $R$ to the carrier space of $(r,s)$. We can also use a
shorthand notation that

\begin{equation*}
Z_{i_{\sigma (1)}}^{i_{1}}\cdots Z_{i_{\sigma (n)}}^{i_{n}}Y_{i_{\sigma
(n+1)}}^{i_{n+1}}\cdots Y_{i_{\sigma (n+m)}}^{i_{n+m}}\,=\mathrm{Tr}[\sigma
Z^{\otimes n}Y^{\otimes m}],
\end{equation*}%
\vspace{1pt}\vspace{1pt}where $\sigma $ is an element of $S_{n+m}.$

The normalization of the restricted Schur polynomial operator is, see for
example \cite{Bhattacharyya:2008rb},
\begin{equation*}
\langle \chi _{R,(r,s)}(Z,Y)\chi _{R,(r,s)}(Z,Y)^{\dagger }\rangle =f_{R}{%
\frac{\mathrm{hooks}_{R}}{\mathrm{hooks}_{r}\,\mathrm{hooks}_{s}}}\,,
\end{equation*}%
where $f_{R}$ is the product of the factors in Young diagram $R$ and $%
\mathrm{hooks}_{R}$ is the product of the hook lengths of Young diagram $R$.
\vspace{1pt}The normalized operators $O_{R,(r,s)}(Z,Y)$ can be obtained by%
\begin{equation}
O_{R,(r,s)\alpha \beta }(Z,Y)=\sqrt{\frac{\mathrm{hooks}_{r}\,\mathrm{hooks}%
_{s}}{f_{R}\,\mathrm{hooks}_{R}}}\chi _{R,(r,s)\alpha \beta }(Z,Y).
\end{equation}

In terms of the normalized operators, the action of the dilatation operator
is, see for example \cite{Koch:2011hb,DeComarmond:2010ie},
\begin{equation*}
DO_{R,(r,s)\alpha \beta }(Z,Y)=\sum_{T,(t,u)}N_{R,(r,s)\alpha \beta
;T,(t,u)\gamma \delta }O_{T,(t,u)\gamma \delta }(Z,Y).
\end{equation*}%
For example at one loop,%
\begin{eqnarray}
N_{R,(r,s)\alpha \beta ;T,(t,u)\gamma \delta } &=&-g_{YM}^{2}\sum_{R^{\prime
}}{\frac{c_{RR^{\prime }}d_{T}nm}{d_{R^{\prime }}d_{t}d_{u}(n+m)}}\sqrt{%
\frac{f_{T}\,\mathrm{hooks}_{T}\,\mathrm{hooks}_{r}\,\mathrm{hooks}_{s}}{%
f_{R}\,\mathrm{hooks}_{R}\,\mathrm{hooks}_{t}\,\mathrm{hooks}_{u}}}\times
\notag \\
&&\times \mathrm{Tr}([\Gamma ^{(R)}(m+1,1),P_{R\rightarrow (r,s)\alpha \beta
}]I_{R^{\prime }~T^{\prime }}[\Gamma ^{(T)}(m+1,1),P_{T\rightarrow
(t,u)\delta \gamma }]I_{T^{\prime }R^{\prime }})\,.  \notag \\
&&  \label{N trace}
\end{eqnarray}
Here the $c_{RR^{\prime }}$ is the weight of the corner box removed from
Young diagram $R$ to obtain Young diagram $R^{\prime }$, and similarly $%
T^{\prime }$ is a Young diagram obtained from $T$ by removing a box. The $%
d_{u}$ denotes the dimension of symmetric group irrep $u$. The intertwiner
operator $I_{R_{1}T_{1}}$ is a map from the carrier space of irreducible
representation $R_{1}$ to the carrier space of irreducible representation $%
T_{1}$. The $I_{R_{1}T_{1}}$ is non-zero if $R_{1}$ and $T_{1}$ are Young
diagrams of the same shape. For the operators with $p$ long rows in the
Young diagram $R$, we remove $m_{i}$ impurities from each $i$-th row to
obtain the Young diagram $r$, and we may denote $\{m_{i}|_{i=1,...,p}\}$ as $%
\vec{m}.~$The $p=2~$case is relatively the most elementary situation in the
discussions here.

After performing the trace we have
\begin{equation}
DO_{R,(r,s)\alpha \beta }=-g_{YM}^{2}\sum_{u,\gamma \delta }\sum_{1\leq
i<j\leq p}M_{s\alpha \beta ;u\gamma \delta }^{(ij)}\Delta
_{ij}O_{R,(r,u)\gamma \delta }
\end{equation}%
where $\Delta _{ij}$ acts only on the Young diagrams $R,r$. The $M_{s\alpha
\beta ;u\gamma \delta }^{(ij)}$ is a mixing matrix in the space of the Young
diagrams of impurities. The action of the operator $\Delta _{ij}$ can be
written as
\begin{equation}
\Delta _{ij}=\Delta _{ij}^{+}+\Delta _{ij}^{0}+\Delta _{ij}^{-}.
\label{Delta_action_terms}
\end{equation}%
We denote the length of the $i$-th row of $r$ by $r_{i}$. The Young diagram $%
r_{ij}^{+}$ is obtained by removing a box from row $j$ and then adding it to
row $i$. The Young diagram $r_{ij}^{-}$ is obtained by removing a box from
row $i$ and then adding it to row $j$. In terms of these Young diagrams we
have that%
\begin{equation*}
\Delta _{ij}^{0}O_{R,(r,s)\alpha \beta }=-(2N+r_{i}+r_{j})O_{R,(r,s)\alpha
\beta },
\end{equation*}%
\begin{equation}
\Delta _{ij}^{+}O_{R,(r,s)\alpha \beta }=\sqrt{(N+r_{i})(N+r_{j})}%
O_{R_{ij}^{+},(r_{ij}^{+},s)\alpha \beta },  \label{Delta_action}
\end{equation}%
\begin{equation*}
\Delta _{ij}^{-}O_{R,(r,s)\alpha \beta }=\sqrt{(N+r_{i})(N+r_{j})}%
O_{R_{ij}^{-},(r_{ij}^{-},s)\alpha \beta }.
\end{equation*}%
The $\Delta _{ij}$ acts on $R,r$ and on the $Z^{\prime }$s. Note that the $R$
and $r$ change in the same way. The $c_{i}=N+r_{i}$ is the factor of the
corner box in the $i$-th row, while $c_{j}=N+r_{j}~$is the factor of the
corner box in the $j$-th row, and they are of order $N$.~The number of rows
in the Young diagrams $R,r$ is $p$, and~the length of the $p$ rows are long
and of order $N$. The Young diagram $s$ of impurities has no more than $p$
rows. The $\Delta _{ij}$ acts on each pair of rows $(i,j).~$For these
operators with a very large dimension of order $O(N)$, the nonplanar
diagrams already contribute in the leading order, and the spectra are
obtained by summing over planar and nonplanar Feynman diagrams.~

These expressions are evaluated in a large $N$ but nonplanar limit. The
expressions (\ref{Delta_action}) are written in the case for the AdS giants.
The expressions for the sphere giants are similar. The weights are now $%
c_{i}=N-r_{i},c_{j}=N-r_{j},$ corresponding to the factors of the corner
boxes on the $i$-th and $j$-th column.$~$In other words, we replace $%
N+r_{i}~ $by $N-r_{i},~$and $N+r_{j}~$by $N-r_{j}$ in the above equations
for the actions of $\Delta _{ij}^{0},\Delta _{ij}^{+}$ and $\Delta _{ij}^{-}$%
.

The construction of the restricted Schur polynomial operators in various
sectors, such as $su(2)$, $su(2|3),$ $sl(2)$ and $su(3)$ sectors, and their
anomalous dimensions have been considered in the recent work for example
\cite{Koch:2011hb,DeComarmond:2010ie,Carlson:2011hy,deMelloKoch:2011ci,
deMelloKoch:2012ck,deMelloKoch:2012sv,deMelloKoch:2011vn,
Koch:2010gp,Koch:2012sf,Koch:2011jk,Koch:2013xaa}. There are many other very
interesting works on giant graviton excitations from various different
perspectives, for example \cite%
{Berenstein:2006qk,Balasubramanian:2004nb,deMelloKoch:2007uu}.

\vspace{1pt}

\vspace{1pt}


\section{$su(2)$ algebra and oscillator algebra}

\label{sec: su(2) contraction} 

The $p$ number of giant graviton D3-branes are expected to have an $u(p)$
symmetry. The $u(p)$ symmetry algebra, with $p\geq 2$, contains the $u(2)$
algebra as a subalgebra, which in turn, contains the $su(2)$ algebra as a
subalgebra. The $u(2)$ algebra is the symmetry algebra acting on a pair of
giant gravitons. The $su(2)$ algebra is embedded as
\begin{equation}
su(2)\subset u(2)\subseteq u(p).
\end{equation}

We first review the construction of the $u(2)$ algebra from the $u(p)$
algebra that was performed in \cite{deMelloKoch:2011ci}. The fundamental
representation of the $u(p)$ algebra represents the elements of the Lie
algebra as $p\times p$ matrices. The generators $E_{ik}\in u(p)~$can be
written as
\begin{equation*}
(E_{ik})_{ab}=\delta _{ia}\delta _{kb},\qquad 1\leq i,k,a,b\leq p\,.
\end{equation*}

The $p$ operators $E_{ii}$ commute with each other so we can choose a basis
in which they are diagonal at the same time. The restricted Schur polynomial
labelled by the Young diagrams is identified with the state with $\{E_{ii}\}$%
. The Young diagrams $r$ with $p$ rows are the irreducible representations
of the symmetric group and also the irreducible representations of the $u(p)$
group. There is a map $\frac{1}{2}E_{ii}\mapsto c_{i},~$for $i=1,...,p,~$%
where $c_{i}$ are the factors of the corner boxes of each row $i$. The \{$%
\frac{1}{2}E_{ii}|i=1,...,p$\} correspond to the $\{c_{i}|i=1,...,p\}.~$The
description for the case of $p$ long columns is similar, where the $c_{i}$
are the factors of the corner boxes of each column $i$, for $i=1,...,p.~$The
$p=2$ case is relatively the simplest case in this construction.

We consider the generators
\begin{equation*}
Q_{ij}={\frac{1}{2}(E_{ii}-E_{jj})},\qquad Q_{ij}^{+}=E_{ij},\qquad
Q_{ij}^{-}=E_{ji}\,,
\end{equation*}%
which obey the $su(2)$ algebra of angular momentum raising and lowering
operators \cite{deMelloKoch:2011ci},
\begin{equation}
\lbrack Q_{ij},Q_{ij}^{+}]=Q_{ij}^{+},\qquad \lbrack
Q_{ij},Q_{ij}^{-}]=-Q_{ij}^{-},\qquad \lbrack
Q_{ij}^{+},Q_{ij}^{-}]=2Q_{ij}\,.
\end{equation}
We can also define
\begin{equation*}
\qquad Q_{ij}^{+}=Q_{ij}^{1}+iQ_{ij}^{2},\qquad
Q_{ij}^{-}=Q_{ij}^{1}-iQ_{ij}^{2},~~~~\ ~~Q_{ij}=Q_{ij}^{3},
\end{equation*}%
and $\mathrm{span}$\{$Q_{ij}^{1},Q_{ij}^{2},Q_{ij}^{3}$\} is the $su(2)$
algebra.

The representations of these $su(2)$ subalgebras can be labelled with the
eigenvalue of ${\vec{Q}}_{ij}^{2}=\frac{1}{2}%
(Q_{ij}^{+}Q_{ij}^{-}+Q_{ij}^{-}Q_{ij}^{+})+(Q_{ij})^{2}~$and the eigenvalue
of $Q_{ij}=Q_{ij}^{3}.~$The $\eta $ and $\Lambda (\Lambda +1)$ are defined
as the eigenvalues of the operators $Q_{ij}$ and ${\vec{Q}}_{ij}^{2}$
respectively. The states are labelled by $|\eta ,\Lambda \rangle$ and
\begin{equation*}
Q_{ij}^{+}|\eta ,\Lambda \rangle =\sqrt{(\Lambda +\eta +1)(\Lambda -\eta )}%
|\eta +1,\Lambda \rangle \,,\qquad
\end{equation*}%
\begin{equation*}
Q_{ij}^{-}|\eta ,\Lambda \rangle =\sqrt{(\Lambda +\eta )(\Lambda -\eta +1)}%
|\eta -1,\Lambda \rangle \,,\qquad
\end{equation*}%
where $-\Lambda \leq \eta \leq \Lambda \,.~$We have so far reviewed the
construction of the $su(2)$ algebra \{$Q_{ij}^{1},Q_{ij}^{2},Q_{ij}^{3}$%
\}~from the $u(p)$ algebra performed in \cite{deMelloKoch:2011ci}.

Let's now turn to the action of the dilatation operator $\Delta _{ij}$ on
these Young diagram operators. We focus on a pair of giant gravitons
labelled by $i$ and $j$, which also correspond to a pair of long rows
labelled by $i$ and $j$. In particular, the operators $\Delta _{ij}$ are, as
according to Eqs. (\ref{Delta_action_terms}) and (\ref{Delta_action}), for
example \cite{Koch:2011hb,Carlson:2011hy,deMelloKoch:2012ck},
\begin{equation}
\Delta _{ij}=-{\frac{1}{2}}(E_{ii}+E_{jj})+Q_{ij}^{-}+Q_{ij}^{+},
\end{equation}
so we can write it in the form
\begin{equation}
\Delta _{ij}=2A_{ij}^{3}-(c_{i}+c_{j})I_{ij},
\end{equation}
where we also define an abelian generator $I_{ij}={\frac{1}{2(c_{i}+c_{j})}}%
(E_{ii}+E_{jj}).~$The eigenvalue of the $Q_{ij}$ is $\frac{1}{2}%
(c_{i}-c_{j}) $~and the $\Lambda =\frac{1}{2}\max |c_{i}-c_{j}|.$

We can define a different set of $su(2)$ generators which will be convenient
for the evaluation of the $\Delta _{ij}$. These can be defined as
\begin{eqnarray}
A^{+} &=&{\frac{1}{2}}(E_{ii}-E_{jj})+{\frac{1}{2}(}%
E_{ij}-E_{ji})=Q_{ij}^{3}+iQ_{ij}^{2},  \notag \\
A^{-} &=&{\frac{1}{2}}(E_{ii}-E_{jj})-{\frac{1}{2}(}%
E_{ij}-E_{ji})=Q_{ij}^{3}-iQ_{ij}^{2}, \\
A^{3} &=&{\frac{1}{2}(}E_{ij}+E_{ji})={\frac{1}{2}(}%
Q_{ij}^{+}+Q_{ij}^{-})=Q_{ij}^{1}.  \notag
\end{eqnarray}
The algebra
\begin{equation*}
\mathrm{span}\text{\{}A^{+},A^{-},A^{3}\text{\}}
\end{equation*}
is the $su(2)$ algebra with commutation relations
\begin{equation}
[A^{-},A^{+}]=2A^{3},~~[A^{3},A^{+}]=-A^{+},~~[A^{3},A^{-}]=A^{-}.
\end{equation}
It is transformed from $\mathrm{span}$\{$Q_{ij}^{1},Q_{ij}^{2},Q_{ij}^{3}$%
\}~by an automorphism of $su(2)$. In this paper, we mainly work with the new
basis \{$A^{+},A^{-},A^{3}$\}.

The linear span of
\begin{equation}
\mathrm{span}\text{\{}A^{+},A^{-},A^{3},I\text{\}}
\end{equation}
forms the $su(2)\times u(1)$ algebra, where $I$ is the generator of the $u(1)
$ algebra that we also include. The $su(2)\times u(1)$ algebra is also the $%
u(2)$ algebra. We now denote $\Delta _{ij}$ as $\Delta _{(1)ij}$. We can
conveniently express
\begin{equation*}
\Delta_{(1)ij}=2A^{3}-(c_{i}+c_{j})I
\end{equation*}
where we have suppressed the $ij$ indices. In particular $\Delta _{(1)ij}$
is a linear combination of $A^{3}$ and $I$. This expression is according to
the findings in for example \cite%
{Koch:2011hb,Carlson:2011hy,deMelloKoch:2012ck}.

We can perform a Inonu-Wigner contraction of this $su(2)\times u(1)$
algebra, by a linear transformation
\begin{equation}
\left[
\begin{array}{c}
a^{\dagger } \\
a \\
\frac{1}{2}\Delta _{(1)ij} \\
1%
\end{array}%
\right] =\left[
\begin{array}{cccc}
\xi & 0 & 0 & 0 \\
0 & \xi & 0 & 0 \\
0 & 0 & 1 & -\frac{1}{2}\xi ^{-2} \\
0 & 0 & 0 & 1%
\end{array}%
\right] \left[
\begin{array}{c}
A^{+} \\
A^{-} \\
A^{3} \\
I%
\end{array}%
\right] ,
\end{equation}
where $\xi =\frac{1}{\sqrt{c_{i}+c_{j}}}$ and is of order $O(\frac{1}{\sqrt{N%
}})$. The inverse transformation is
\begin{equation}
\left[
\begin{array}{c}
A^{+} \\
A^{-} \\
A^{3} \\
I%
\end{array}%
\right] =\left[
\begin{array}{cccc}
\xi ^{-1} & 0 & 0 & 0 \\
0 & \xi ^{-1} & 0 & 0 \\
0 & 0 & 1 & \frac{1}{2}\xi ^{-2} \\
0 & 0 & 0 & 1%
\end{array}%
\right] \left[
\begin{array}{c}
a^{\dagger } \\
a \\
\frac{1}{2}\Delta _{(1)ij} \\
1%
\end{array}
\right] .
\end{equation}
This Inonu-Wigner contraction corresponds to the limit $\xi ^{2}\Delta
_{(1)ij}\ll 1$,~in other words the eigenvalue of $\frac{1}{2}\Delta _{(1)ij}$%
,~which is the oscillator level, is much smaller than $N$. The above $%
su(2)\times u(1)$ algebra, which is also an $u(2)$ algebra, via the
Inonu-Wigner contraction, becomes the harmonic oscillator algebra,
\begin{equation}
\mathrm{span}\text{\{}a^{\dagger },a,\Delta _{(1)ij},1\text{\}}
\end{equation}
with commutation relations
\begin{equation}
[a,a^{\dagger }]=1,\ \ [\Delta _{(1)ij},a^{\dagger }]=-2a^{\dagger
},~~[\Delta _{(1)ij},a]=2a.
\end{equation}
This is a harmonic oscillator algebra obeyed by the creation and
annihilation operators of this oscillator.

The $u(p)$ symmetry algebra always includes $u(2)$ symmetry algebra as a
subalgebra. From the point of view of the operators given by Young diagrams,
the $u(2)$ symmetry algebra is the symmetry algebra acting on two long rows,
the $i$-th row and the $j$-th row. From the point of view of the gravity
side, the $u(2)$ symmetry algebra is the symmetry algebra of two giant
graviton branes labelled by $i$ and $j$.

The existence of $u(p)$ symmetry is also from the Young diagrams with $p$
long rows, which also label the irreducible representations of the symmetric
algebras \cite{Koch:2011hb}. This involves the duality between
representations of symmetric groups and unitary groups that we have
discussed.

The $u(p)$ symmetry can also be observed from the symmetry of $p$ giant
gravitons. The $u(p)$ symmetry can also be seen from the bubbling geometries
in \cite{Lin:2004nb} with a white disk of area $p$ inside of the black disk
of area $N$. The white disk with area $p$ is the geometric dual of $p$ giant
gravitons. If we zoom in near the white disk it approaches another AdS space
and is dual to $p$ number of three-branes with $u(p)$ symmetry.


\section{Higher loop anomalous dimensions and oscillator algebra}

\label{sec: oscillator algebra}

We now discuss relation between the oscillator algebra and higher loop
dilatation operators. The anomalous dimension $\gamma (g)$ expanded at one
and two loops is the eigenvalue of
\begin{equation*}
\hat{D}=\hat{D}_{2}+\hat{D}_{4}
\end{equation*}
with the one loop dilatation operator
\begin{equation}
\hat{D}_{2}=-2g:\mathrm{Tr}\left( \left[ Z,Y\right] \left[ \partial
_{Z},\partial _{Y}\right] \right) :
\end{equation}
and the two loop dilatation operator \cite{Beisert:2003tq}
\begin{eqnarray}
\hat{D}_{4} &=&-2g^{2}:\mathrm{Tr}\left( \left[ \left[ Z,Y\right] ,\partial
_{Z}\right] \left[ \left[ \partial _{Z},\partial _{Y}\right] ,Z\right]
\right) :-2g^{2}:\mathrm{Tr}\left( \left[ \left[ Z,Y\right] ,\partial _{Y}%
\right] \left[ \left[ \partial _{Z},\partial _{Y}\right] ,Y\right] \right) :
\notag \\
&&-2g^{2}:\mathrm{Tr}\left( \left[ \left[ Z,Y\right] ,T^{a}\right] \left[ %
\left[ \partial _{Z},\partial _{Y}\right] ,T^{a}\right] \right) :
\end{eqnarray}
where in the convention of \cite{Beisert:2003tq} $g={\frac{{\tilde{g}}%
_{YM}^{2}}{16\pi ^{2}}}$.{~}The normalization for the $\hat{D}_{2}$ and $%
\hat{D}_{4}$ is in the convention of \cite{Beisert:2003tq}.{~}The
normalization in this convention for $\hat{D}_{2}$ and $\hat{D}_{4}$ is a
factor of two larger than the normalization used in the convention of \cite%
{Koch:2011hb,deMelloKoch:2012ck,Carlson:2011hy,
deMelloKoch:2011vn,DeComarmond:2010ie,Koch:2011jk,deMelloKoch:2011ci}, for
example. Here we denote $D_{2}$ and $D_{4}$ for the convention in \cite%
{Koch:2011hb,deMelloKoch:2012ck,Carlson:2011hy,
deMelloKoch:2011vn,DeComarmond:2010ie,Koch:2011jk,deMelloKoch:2011ci}.

The action of the two loop dilatation operator on the restricted Schur
polynomials has been evaluated by \cite{deMelloKoch:2012sv}, and it is given
by,
\begin{equation}
\hat{D}_{4}O_{R,(r,s)\alpha \beta }=-2g^{2}\sum_{u,\,\gamma \delta
}\sum_{i<j}M_{s\alpha \beta ;u\gamma \delta }^{(ij)}\Delta
_{(2)ij}O_{R,(r,u)\gamma \delta }
\end{equation}
where
\begin{eqnarray*}
M_{s\alpha \beta ;u\gamma \delta }^{(ij)} &=&{\frac{m}{\sqrt{d_{s}d_{u}}}}%
\left( \left\langle \vec{m},s,\beta \,;\,a|E_{ii}^{(1)}|\vec{m},u,\delta
\,;\,b\right\rangle \left\langle \vec{m},u,\gamma \,;\,b|E_{jj}^{(1)}|\vec{m}%
,s,\alpha \,;\,a\right\rangle \right. \\
&&+\left. \left\langle \vec{m},s,\beta \,;\,a|E_{jj}^{(1)}|\vec{m},u,\delta
\,;\,b\right\rangle \left\langle \vec{m},u,\gamma \,;\,b|E_{ii}^{(1)}|\vec{m}%
,s,\alpha \,;\,a\right\rangle \right)
\end{eqnarray*}%
in which $a$ and $b$ are summed. The $a$ labels states in irreducible
representation $s$ and the $b$ labels states in irreducible representation $%
u $.

We denote $\Delta _{(1)ij}=\Delta _{ij}$ for the one loop dilatation
operator and $\Delta _{(2)ij}$ for the two loop dilatation operator. The $%
\Delta _{(2)ij}$ has been computed by \cite{deMelloKoch:2012sv}, and it can
be written as a sum of two terms. We have defined that
\begin{equation*}
\Delta _{ij}^{0}O_{R,(r,s)\alpha \beta }=-(2N+r_{i}+r_{j})O_{R,(r,s)\alpha
\beta },
\end{equation*}
\begin{equation}
\Delta _{ij}^{+}O_{R,(r,s)\alpha \beta }=\sqrt{(N+r_{i})(N+r_{j})}%
O_{R_{ij}^{+},(r_{ij}^{+},s)\alpha \beta },
\end{equation}
\begin{equation*}
\Delta _{ij}^{-}O_{R,(r,s)\alpha \beta }=\sqrt{(N+r_{i})(N+r_{j})}%
O_{R_{ij}^{-},(r_{ij}^{-},s)\alpha \beta }.
\end{equation*}
The $\Delta _{(2)ij}=\Delta _{ij}^{(1)}+\Delta _{ij}^{(2)}$ and
can be written as
\begin{equation}
\Delta _{ij}^{(1)}=n(\Delta _{ij}^{+}+\Delta _{ij}^{0}+\Delta _{ij}^{-}),
\end{equation}%
\begin{equation}
\Delta _{ij}^{(2)}=(\Delta _{ij}^{+})^{2}+\Delta _{ij}^{0}\Delta
_{ij}^{+}+2\Delta _{ij}^{+}\Delta _{ij}^{-}+\Delta _{ij}^{0}\Delta
_{ij}^{-}+(\Delta _{ij}^{-})^{2}.
\end{equation}%
As the same as $\Delta _{(1)ij},~$the $\Delta _{(2)ij}$ acts on each pair of
rows $(i,j).~$The $c_{i},c_{j}$ and $n$ are of order $N.$

We can simplify the two-loop dilatation operator $\Delta _{(2)ij} $ as
\begin{equation}
\Delta _{(2)ij}=(\Delta _{(1)ij}+(c_{i}+c_{j}+n)I)\Delta _{(1)ij}
\end{equation}
when acting on these Young diagram operators. These expressions are
evaluated in the large $N$ but nonplanar limit. From this expression, we see
that the operators that are eigenstates of $\Delta _{(1)ij}$, are also
eigenstates of $\Delta _{(2)ij}$. Since $\Delta _{(1)ij}$ mixes the
operators that differ by moving at most one box in the ($R,r$)~irreducible
representations, $\Delta _{(2)ij}$ mixes the operators that differ by moving
at most two boxes in the ($R,r$)~irreducible representations.~The action of $%
(\Delta _{ij}^{+})^{h},(\Delta _{ij}^{-})^{h}$ mix the operators that differ
by moving at most $h$ boxes in the ($R,r$)~irreducible representations, and
therefore $(\Delta _{(1)ij})^{h}$ mixes the operators that differ by moving
at most $h$ boxes in the ($R,r$)~irreducible representations.

The eigenstates in the oscillator basis are states of finite harmonic
oscillators, see for example \cite%
{Koch:2011hb,Carlson:2011hy,deMelloKoch:2012ck}. These operators can be
written as
\begin{equation}
O_{q_{ij}}(\sigma )=\sum\limits_{R,r}\tilde{f}_{q_{ij}}^{R,r}O_{R,r}(\sigma
)=\sum\limits_{R,r}\sum_{s,\alpha \beta }\tilde{f}_{q_{ij}}^{R,r}C_{\alpha
\beta }^{s}(\sigma )O_{R,(r,s)\alpha \beta }
\end{equation}%
where$~\tilde{f}_{q_{ij}}^{R,r}$ are the wave functions of the discrete
harmonic oscillator. The Young diagrams $(R,r)~$both have $p$ long rows, and
$R\vdash n+m,~r\vdash n$. Those functions also appear in the study of models
of finite harmonic oscillators, for example \cite{finiteoscillator}. The $%
C_{\alpha \beta }^{s}(\sigma )$ are group theoretic coefficients and $%
O_{R,r}(\sigma )$ are the operators labelled by a permutation graph $\sigma$
that maps \cite{deMelloKoch:2012ck} to the data $\{n_{ij}\}$ via the
function $n_{ij}(\sigma )$, where $n_{ij}$~are the number of$~$strings
stretching between the two branes labelled by $i$ and $j.$

The eigenvalues of $\Delta _{ij}$ acting on $O_{q_{ij}}(\sigma )~ $is $%
4q_{ij},$ where $q_{ij}$ is an integer \cite{Koch:2011hb,Carlson:2011hy}
denoting the level of the harmonic oscillator for the pair of giant
gravitons $i$ and $j$, that is,%
\begin{equation}
\Delta _{(1)ij}O_{q_{ij}}(\sigma )=4q_{ij}O_{q_{ij}}(\sigma ),
\label{Delta_1 eigen}
\end{equation}%
and%
\begin{equation}
\hat{D}_{2}O_{q_{ij}}(\sigma )=-2g\sum\limits_{R,r}\tilde{f}%
_{q_{ij}}^{R,r}n_{ij}\Delta _{ij}O_{R,r}(\sigma
)=-8gq_{ij}n_{ij}O_{q_{ij}}(\sigma ).
\end{equation}%
The $q_{ij}~$is bounded above due to the discrete finite harmonic
oscillator. Therefore we consider the simple case in which $q_{ij}\ll N$. We
see that
\begin{equation}
\Delta _{(2)ij}O_{q_{ij}}(\sigma )=4q_{ij}(c_{i}+c_{j}+n)O_{q_{ij}}(\sigma
)=(c_{i}+c_{j}+n)\Delta _{(1)ij}O_{q_{ij}}(\sigma )  \label{Delta_2 Delta_1}
\end{equation}%
in the large $N$ limit, and
\begin{equation}
\hat{D}_{4}O_{q_{ij}}(\sigma )=-2g^{2}\sum\limits_{R,r}\tilde{f}%
_{q_{ij}}^{R,r}n_{ij}\Delta _{(2)ij}O_{R,r}(\sigma
)=-8g^{2}q_{ij}n_{ij}(c_{i}+c_{j}+n)O_{q_{ij}}(\sigma ).
\end{equation}

In this case, the anomalous dimension $\gamma (\lambda )$ expanded at one
and two loop orders is
\begin{equation}
\gamma _{(1)}={\frac{8q_{ij}n_{ij}}{N}}\lambda ,~~~~~~~~~~~~\gamma _{(2)}={%
\frac{8q_{ij}n_{ij}}{N}}({\frac{c_{i}+c_{j}+n}{N}})\lambda ^{2},
\end{equation}%
where ${\frac{c_{i}+c_{j}+n}{N}=\frac{r_{i}+r_{j}+2N+n}{N}}$~is of order 1{,}%
${~}q_{ij}=0,1,2,..,q_{\max }$,$~$and the $\lambda $ denotes $gN$.~This is
the simple case that there are equal number of strings emanating from brane $%
i$ to brane $j$, and from brane $j$ to brane $i$, and these numbers are $%
n_{ij}^{+}$ and $n_{ij}^{-}$ respectively. In this case$%
~n_{ij}=n_{ij}^{+}+n_{ij}^{-}=2n_{ij}^{+}$, where $n_{ij}^{+}~$is a
non-negative integer. If $|c_{i}-c_{j}|\ll c_{i}+c_{j},~$then $r_{i}+r_{j}$
is approximately $2r_{j}~$\cite{deMelloKoch:2012sv}. Note that in this paper
we also assumed the special case that we are looking at the operators whose $%
q_{\max }$ is much smaller than $N$.

From Eq. (\ref{Delta_2 Delta_1}), the commutation relations of $\Delta
_{(2)ij}$ are
\begin{equation}
\lbrack a,a^{\dagger }]=1,\ \ [\Delta _{(2)ij},a^{\dagger
}]=-2(c_{i}+c_{j}+n)a^{\dagger },~~[\Delta _{(2)ij},a]=2(c_{i}+c_{j}+n)a.
\end{equation}%
So the two-loop dilatation operator $\Delta _{(2)ij}$ also satisfies the
oscillator algebra. This relation was observed in \cite{deMelloKoch:2012sv}.

We can write them in terms of polynomials of $A^{3}$. Using the relation
\begin{equation}
\Delta _{(1)ij}=2A^{3}-(c_{i}+c_{j})I
\end{equation}%
the $\Delta _{(2)ij}$ can be rewritten as%
\begin{eqnarray}
\Delta _{(2)ij} &=&(2A^{3}-(c_{i}+c_{j})I)(2A^{3}+nI)  \notag \\
&=&4(A^{3})^{2}+2(n-c_{i}-c_{j})A^{3}-n(c_{i}+c_{j})I.
\end{eqnarray}
So $\Delta _{(2)ij}$ is a polynomial of $A^{3}$ of degree 2, that is, $%
P_{2}(A^{3}).$ The $\Delta _{(1)ij}$ is a polynomial of $A^{3}$ of degree 1,
that is, $P_{1}(A^{3})$. By the structure of the dilatation operators at
higher loops, the higher loop dilatation operators $\Delta _{(h)ij}$,~with $%
h $ the loop order, is a polynomial of $A^{3}$ of degree $h$, that is, $%
P_{h}(A^{3}),$
\begin{equation*}
\Delta _{(h)ij}=P_{h}(A^{3})
\end{equation*}
where
\begin{equation}
A^{3}=\frac{1}{2}[\Delta _{(1)ij}+(c_{i}+c_{j})I],
\end{equation}
and they can mix operators that differ by moving at most $h$ boxes in the ($%
R,r$)~irreducible representations in the large $N$ limit that we consider.
These dilatation operators form a polynomial ring $\mathbb{R}[A^{3}]$ over $%
A^{3}$.

The operators with $q_{ij}=0$ should correspond to the BPS states, and they
are the states of the giant gravitons without the non-BPS excitations. These
states are thus the eigenstates of the dilatation operator with the
eigenvalues being zero. Since both $\Delta _{(1)ij}~$and $\Delta _{(2)ij}~$%
contains an overall factor of $(2A^{3}-(c_{i}+c_{j})I)$, this factor acting
on the BPS states is zero. As analyzed above, if $\Delta _{(h)ij}$,~with $h$
the loop order, is a polynomial of $A^{3}$ of degree $h$,~then it is
expected to contain an overall factor of $(2A^{3}-(c_{i}+c_{j})I)$.~It may
be written as
\begin{eqnarray}
\Delta _{(h)ij}
&=&(2A^{3}-(c_{i}+c_{j})I)[\sum_{l=1,...,h}a_{l}(A^{3})^{h-l}]  \notag \\
&=&\Delta _{(1)ij}[N^{h-1}F_{h}+u_{h-2}]  \label{Delta_h}
\end{eqnarray}%
when acting on the operators, where $F_{h}$ is order $O(1)~$coefficient, and
$u_{h-2}\leq O(N^{h-2})$.~The $F_{h}$ is of order $O(1)$ because the
eigenvalue of $(A^{3})^{h-1}$ is of order $O(N^{h-1})$.~In the derivation
from the first line to the second line of Eq. (\ref{Delta_h}), we used that $%
A^{3}O_{q_{ij}}(\sigma )=\frac{1}{2}(c_{i}+c_{j}+\Delta
_{(1)ij})O_{q_{ij}}(\sigma )=[O(N)]O_{q_{ij}}(\sigma ),~$which means that
the eigenvalue of $A^{3}$ on the diagonalized operators is of order $O(N).$
The subleading terms in Eq. (\ref{Delta_h}) are subleading in the large $N$
limit.

So we have the relation
\begin{equation}
\Delta _{(h)ij}O_{q_{ij}}(\sigma )=F_{h}N^{h-1}\Delta
_{(1)ij}O_{q_{ij}}(\sigma )
\end{equation}
in the large $N$ limit. In this relation, we have also assumed that we are
considering the oscillator level to be much smaller than $N$, that is, we
are giving an additional condition $\frac{q_{ij}}{N}\ll 1$ to make
simplifications. The examples of the $h=1$ and $h=2$ are \vspace{1pt}$%
F_{1}=1~$and $F_{2}=\frac{c_{i}+c_{j}+n}{N},~$and are both of order $O(1)$.%
\vspace{1pt} So we have%
\begin{equation*}
\Delta _{(1)ij}O_{q_{ij}}(\sigma )=4q_{ij}O_{q_{ij}}(\sigma )=F_{1}\Delta
_{(1)ij}O_{q_{ij}}(\sigma ),
\end{equation*}%
\begin{equation*}
\Delta _{(2)ij}O_{q_{ij}}(\sigma )=4q_{ij}(c_{i}+c_{j}+n)O_{q_{ij}}(\sigma
)=F_{2}N\Delta _{(1)ij}O_{q_{ij}}(\sigma ).
\end{equation*}%
In the large $N$ limit, from the structure of the dilatation operator,%
\begin{equation}
\Delta _{(h)ij}O_{q_{ij}}(\sigma )=4q_{ij}F_{h}N^{h-1}O_{q_{ij}}(\sigma
)=F_{h}N^{h-1}\Delta _{(1)ij}O_{q_{ij}}(\sigma ),
\end{equation}%
where $F_{h}$ is order $O(1)$.

In the large $N$ limit, as long as the higher loop dilatation operator $%
\Delta _{(h)ij}$ is a polynomial of $A^{3}$,$~$it will satisfy the following
oscillator algebra,%
\begin{equation}
\lbrack a,a^{\dagger }]=1,\ \ [\Delta _{(h)ij},a^{\dagger
}]=-2F_{h}N^{h-1}a^{\dagger },~~[\Delta _{(h)ij},a]=2F_{h}N^{h-1}a.
\label{osc_alg}
\end{equation}%
We refer to Eq. (\ref{osc_alg}) as the oscillator algebra satisfied by $%
\Delta _{(h)ij}$. If the oscillator algebra is satisfied by $\Delta
_{(l)ij},~$with $l=1,...,h$, then the integrability is preserved at the $h$
loop order. The oscillator algebra at $h$-loop means that, among other
aspects, $\Delta _{(h)ij}$ will not change the $q_{ij}$ in Eq. (\ref{Delta_1
eigen}).

\vspace{1pt}Assuming that the oscillator algebra is satisfied in the large $%
N $ limit at all loops, if we sum over $\sum_{h=1}^{\infty }g^{h}\Delta
_{(h)ij}=\Delta $, we have that
\begin{equation*}
\lbrack \sum_{h=1}^{\infty }g^{h}\Delta _{(h)ij},a^{\dagger
}]=\sum_{h=1}^{\infty }g^{h}[\Delta _{(h)ij},a^{\dagger
}]=-2\sum_{h=1}^{\infty }g^{h}F_{h}N^{h-1}a^{\dagger }=-2f(\lambda
)a^{\dagger },
\end{equation*}%
where we define an interpolating function $f(\lambda ),$%
\begin{equation}
\sum_{h=1}^{\infty }g^{h}F_{h}N^{h-1}=\sum_{h=1}^{\infty }\lambda
^{h}F_{h}/N=\frac{1}{N}f(\lambda ).
\end{equation}%
The $f(\lambda )~$is a function of $\lambda ,~$and its coefficients in $%
\lambda $ expansions are also functions of$~\frac{c_{i}}{N},\frac{c_{j}}{N}$.%
$~$Since $F_{1}=1,~$the expansion of $f(\lambda )~$is $f(\lambda )=\lambda
+\sum_{h=2}^{\infty }F_{h}\lambda ^{h}.~$The oscillator algebra satisfied by
the $\Delta $ is hence,
\begin{equation}
[a,a^{\dagger }]=1,\ \ [\Delta ,a^{\dagger }]=-2\frac{f(\lambda )}{N}%
a^{\dagger },~~[\Delta ,a]=2\frac{f(\lambda )}{N}a.
\end{equation}

The dilatation operator at all loops can be expanded as$~\hat{D}%
(g)=\sum_{h=0}^{\infty }\hat{D}_{2h}$,~where $\hat{D}_{2h}$ is at order $%
g^{h}$, with $h$ the loop order. In the large $N$ limit,
\begin{equation*}
g^{h}\Delta _{(h)ij}O_{q_{ij}}(\sigma )=F_{h}g^{h}N^{h-1}\Delta
_{(1)ij}O_{q_{ij}}(\sigma ),
\end{equation*}%
and hence\vspace{1pt} the anomalous dimension is%
\begin{equation}
\gamma =\sum_{h=1}^{\infty }\gamma _{(h)}=8q_{ij}n_{ij}\sum_{h=1}^{\infty
}g^{h}F_{h}N^{h-1}={\frac{8q_{ij}n_{ij}}{N}}f(\lambda ).
\end{equation}%
The one loop anomalous dimension is $\gamma _{(1)}=8{\frac{\lambda }{N}}%
q_{ij}n_{ij}.$ As compared to the one loop expression, the effect at higher
loops is the renormalization of the coupling to an interpolating function,
that is, $\lambda \rightarrow f(\lambda ),$ while keeping the same $%
q_{ij}n_{ij}$~dependence which is protected by the oscillator algebra. The $%
f(\lambda )$ is also a function of $\frac{c_{i}}{N},\frac{c_{j}}{N},~$and
may be written also as $f(\lambda ;\frac{c_{i}}{N},\frac{c_{j}}{N}).$

We have focused on the $su(2)$ sector of the operators and on a pair of long
rows labelled by $i$ and $j$ in the Young diagram with total $p$ rows. This
sector has a harmonic oscillator algebra, for which we have presented
evidence that they will persist at higher loops.

The diagonalized operators acted on by the one-loop and two-loop dilatation
operators will not be changed by the $h$-loop dilatation operators if the
oscillator algebra, Eq. (\ref{osc_alg}), is preserved at loop order $h$. So,
if the oscillator algebra is preserved at all loop orders, they would not be
changed at all loops.

Our analysis indicates that the nonplanar integrability is protected in the
large $N$ limit by the oscillator algebra. This is an integrability in a
large $N$ but nonplanar limit. We give further evidence that this
integrability in the nonplanar regime is preserved at higher loops and
possibly at all loops. This is the case if higher loop and all loop
dilatation operators in the large $N$ but nonplanar limit satisfy the
oscillator algebra. The oscillator algebra descends from the $u(2)$ symmetry
algebra inside of $u(p)$ symmetry algebra of the system of $p$ giant
gravitons and such symmetry is expected to be robust under loop corrections.
We have presented evidences that the higher loop dilatation operators in the
large $N$ limit can satisfy the oscillator algebra.

\vspace{1pt}

\vspace{1pt}

\vspace{1pt}


\section{General number of pairs and effective spring constants}

\vspace{1pt}\label{sec: general pairs}

\vspace{1pt}

We now consider pairs of giant gravitons. Each pair of long rows labelled by
$i$ and $j$ correspond to a pair of giant gravitons labelled by $i$ and $j$.
The numbers of strings between each pair of giant gravitons is $n_{ij}$. The
total number of strings is $m$, which is divided into $p$ integers $%
\{m_{i}|i=1,...,p\}.~$The $m_{i}$ denotes the number of strings emanating
from the $i$-th brane. Because of charge conservation on each brane, the
number of strings terminating on the $i$-th brane is also $m_{i}$. The
string configurations are modding out by the permutation symmetry $%
H=S_{m_{1}}\times S_{m_{2}}\times \cdots \times S_{m_{p}}~$on both their
left open-ends and their right open-ends. The open string configurations are
in one-to-one correspondence with elements of the double coset \cite%
{deMelloKoch:2012ck}
\begin{equation}
H_{L}\setminus S_{m}/H_{R}
\end{equation}%
with $m=\sum_{i}m_{i},~$and $H_{L}=H_{R}=H$ which is the subgroup $%
S_{m_{1}}\times S_{m_{2}}\times \cdots \times S_{m_{p}}.$~The $H_{L}$
permutes the $m$ left open-ends and the $H_{R}$ permutes the $m$ right
open-ends. The relation $H_{L}=H_{R}$ is due to the charge conservation on
each brane.

There is a map
\begin{equation}
\sigma \mapsto \{n_{ij}|1\leq i<j\leq p\}
\end{equation}%
where $\sigma ~$is an element of the double coset \cite{deMelloKoch:2012ck}.
The numbers $n_{ij}(\sigma )$ can be given from the element of the double
coset $\sigma $. For a generic configuration of open strings on $p$ giant
gravitons, by distinguishing orientation, the number of strings emanating
from brane $i$ and terminating on brane $j~$is $n_{ij}^{+}$, whereas the
number of strings emanating from brane $j$ and terminating on brane $i$ is $%
n_{ij}^{-}.~$The total number of strings between brane $i$ and $j$, without
distinguishing orientation, is thus $n_{ij}=n_{ij}^{+}+n_{ij}^{-}.~$The
permutation graph $\sigma ~$corresponds to $n_{ij}(\sigma
)=n_{ij}^{+}(\sigma )+n_{ij}^{-}(\sigma )~$strings between the pairs of two
branes labelled by $i~$and $j$, and $m_{i}-\sum_{j\neq i}n_{ij}^{+}=n_{ii}~$%
strings emanating from and terminating on the same brane $i$.~We also have
that $n_{ij}^{+}\leq m_{i}\leq m.$

The operators correspond to the data $\{n_{ij}\}$ and its associated
permutation graph is
\begin{equation}
O_{R,r}(\sigma )=\sum_{s\vdash m}\sum_{\alpha \beta }C_{\alpha \beta
}^{s}(\sigma )O_{R,(r,s)\alpha \beta },
\end{equation}%
where the group theoretic coefficients are \cite{deMelloKoch:2012ck}%
\begin{equation*}
C_{\alpha \beta }^{s}(\sigma )=\frac{\sqrt{d_{s}}|H|}{\sqrt{m!}}%
\sum_{l,j}\Gamma ^{(s)}(\sigma )_{lj}B_{l\alpha }^{s\rightarrow
1_{H}}B_{j\beta }^{s\rightarrow 1_{H}}.
\end{equation*}%
The $B_{l\alpha }^{s\rightarrow 1_{H}}$ are the branching coefficients for
the trivial irrep $1_{H}~$of $H$ inside the representation $s$ of $S_{m}$.
The $B_{l\alpha }^{s\rightarrow 1_{H}}$ gives the expansion of the $\alpha $%
'th occurrence of the identity irrep of $H$ when irrep $s$ of $S_{m}$ is
decomposed into irreps of the subgroup $H$, in terms of the states labelled $%
l$ in $s$. The $l,j$ here are labels of states in the irreducible
representation $s$. Similar methods of representation theory used in this
context were also used in defining the multi-matrix operators in the global
symmetry basis or the flavor basis \cite{Brown:2007xh,Brown:2008ij}.

Since the interaction between the $p$ giant gravitons are pair-wise, the
most elementary situation is the interaction between two giant gravitons
labelled by $i$ and $j$, where $1\leq i<j\leq p$.~A simple situation is
provided by a pair of giant gravitons with equal number of strings from each
other, that is $n_{ij}=n_{ij}^{+}+n_{ij}^{-}=2n_{ij}^{+}$. This also
corresponds to a pair of long rows labelled by $i$ and $j$.

For $p=2$, which is the simplest and the most elementary case, $H$ is $%
S_{m_{1}}\times S_{m_{2}}.$ The $m_{1}+m_{2}$ left open-ends of the strings
are divided into $m_{1}$ and $m_{2}~$of them, respectively, on the two giant
gravitons. Because of charge conservation, the number of right open-ends are
also $m_{1}$ and $m_{2}$, respectively, on the two giant gravitons. The left
open-ends and the right open-ends are related by a permutation$~$in $%
S_{m_{1}+m_{2}}$.~The open string configurations are thus in one-to-one
correspondence with elements of the double coset
\begin{equation}
H\setminus S_{m_{1}+m_{2}}/H,
\end{equation}
since the left coset corresponds to modding out by the $S_{m_{1}}\times
S_{m_{2}}$ on the left open-ends of the strings, and the right coset
corresponds to modding out by the $S_{m_{1}}\times S_{m_{2}}$ on the right
open-ends of the strings. For the simplest case that $p=2$, the permutation
graph corresponds to $n_{12}=2n_{12}^{+}=2n_{12}^{-}~$strings between the
two branes labelled by $1~$and $2$, $m_{1}-n_{12}^{+}~$strings emanating and
terminating on the same brane $1$, and $m_{2}-n_{12}^{-}$ strings emanating
and terminating on the same brane $2$.

The action of the dilatation operator on the operator$~O_{R,r}(\sigma )$ is
\begin{equation}
DO_{R,r}(\sigma )=-g_{YM}^{2}\sum_{1\leq i<j\leq p}n_{ij}(\sigma )\Delta
_{ij}O_{R,r}(\sigma ).
\end{equation}%
The $p$ long rows correspond to the $p$ giant gravitons.~The $n_{ij}(\sigma
) $ is a map $\sigma \mapsto \{n_{ij}\}.~$The excitation energy of the
system of giant gravitons $\delta E$ is given by the eigenvalue of the
operator $\sum g_{YM}^{2}n_{ij}\Delta _{ij}.~$

Since the operator $O_{q_{ij}}(\{n_{ij}\})$ in the oscillator basis is
\begin{equation}
O_{q_{ij}}(\{n_{ij}\})=O_{q_{ij}}(\sigma )=\sum\limits_{R,r}\tilde{f}%
_{q_{ij}}^{R,r}O_{R,r}(\sigma ),
\end{equation}%
and
\begin{equation*}
\Delta _{ij}O_{q_{ij}}(\sigma )=4q_{ij}O_{q_{ij}}(\sigma ),
\end{equation*}%
where $q_{ij}$ is a non-negative integer, we have that
\begin{equation}
D_{2}O_{q_{ij}}(\sigma )=-g_{YM}^{2}\sum\limits_{R,r}\tilde{f}%
_{q_{ij}}^{R,r}n_{ij}(\sigma )\Delta _{ij}O_{R,r}(\sigma
)=-4g_{YM}^{2}q_{ij}n_{ij}O_{q_{ij}}(\sigma ).
\end{equation}%
The action of the dilatation operator $\hat{D}_{4}$ on these operators is
\begin{equation}
\hat{D}_{4}O_{R,r}(\sigma )=-2g^{2}\sum_{i<j}n_{ij}(\sigma )\Delta
_{(2)ij}\,O_{R,r}(\sigma )
\end{equation}%
\vspace{1pt}which was obtained in \cite{deMelloKoch:2012sv}, and%
\begin{equation}
\hat{D}_{4}O_{q_{ij}}(\sigma )=-2g^{2}\sum\limits_{R,r}\tilde{f}%
_{q_{ij}}^{R,r}n_{ij}(\sigma )\Delta _{(2)ij}O_{R,r}(\sigma
)=-8g^{2}q_{ij}n_{ij}(c_{i}+c_{j}+n)O_{q_{ij}}(\sigma ).
\end{equation}

We focus on a pair of branes labelled by $i$ and $j$. We consider the simple
case that there are equal numbers of strings emanating from brane $i$ to
brane $j$, and from brane $j$ to brane $i$. By charge conservation on each
brane, in this case,$~n_{ij}^{+}=n_{ij}^{-}$ and $%
n_{ij}=2n_{ij}^{+}=2n_{ij}^{-}$. The action of the dilatation operator is%
\begin{equation}
DO_{R,r}(\sigma )=-g_{YM}^{2}n_{ij}(\sigma )\Delta _{ij}O_{R,r}(\sigma ).
\end{equation}%
The spectrum with general number of pairs of giant gravitons has been
studied in \cite{deMelloKoch:2011ci}. Note that the notation $2n_{ij}$ used
in \cite{deMelloKoch:2011ci} is the $2n_{ij}^{+}=n_{ij}$ in the discussion
here. The frequency $2g_{YM}^{2}n_{ij}=4g_{YM}^{2}n_{ij}^{+}$ is analogous
\cite{deMelloKoch:2011ci} to the spring constant $k_{ij}$ of the pair of
giant gravitons $(i,j),$%
\begin{equation}
k_{ij}=4n_{ij}^{+}\frac{\lambda }{N}.
\end{equation}%
We see that $k_{ij}$ is a function of the coupling constant, that is $%
k_{ij}=k_{ij}(\lambda )$. The excitation energy $\delta E$ of the system of
the giant gravitons is
\begin{equation}
\delta E=2q_{ij}k_{ij}=8g_{YM}^{2}q_{ij}n_{ij}^{+}=8\frac{\lambda }{N}%
q_{ij}n_{ij}^{+}.
\end{equation}
Here we also recover the $p=2$ case \cite{Carlson:2011hy,Koch:2011hb}. Our
convention here is $\delta E={\hat{\gamma}=~}\frac{1}{2}\gamma ,$ where
normalization convention for ${\hat{\gamma}}$ was used in \cite%
{Koch:2011hb,deMelloKoch:2012ck,Carlson:2011hy,
deMelloKoch:2011vn,DeComarmond:2010ie,Koch:2011jk,deMelloKoch:2011ci} and
normalization convention for $\gamma $ was used in \cite{Beisert:2003tq}.

Our analysis suggests that the higher loop dilatation operator in the large $%
N$ limit will satisfy the oscillator algebra. So they will not change the $%
q_{ij}n_{ij}$ dependence of the spectra. We have provided evidence for this
from the oscillator algebra. They will renormalize the effective coupling
and the coefficient $\frac{\lambda }{N}$ to $\frac{f(\lambda )}{N}.$ The $%
f(\lambda )$ is an effective coupling and also an interpolating function. So
the effective spring constants become $k_{ij}=4n_{ij}^{+}\frac{f(\lambda )}{N%
}.~$In the derivation of this, we have also assumed that we are considering
the oscillator level to be much smaller than $N$, in other words $\frac{%
q_{ij}}{N}\ll 1$. According to the renormalization of the effective coupling
constant at higher loops, we have
\begin{equation}
\delta E=2q_{ij}k_{ij}=8\frac{f(\lambda )}{N}q_{ij}n_{ij}^{+}.
\end{equation}
The effective Hamiltonian of this quantum mechanic system is a system of
masses with kinetic energies and with pairwise potentials proportional to $%
k_{ij}|x_{i}-x_{j}|^{2}~$which are quadratic functions of the relative
displacements between pairs of such masses. In other words, according to our
analysis, the spring constants will also be renormalized.

\vspace{1pt}

\vspace{1pt}

\vspace{1pt}


\section{Discussion}

\label{sec: discussion}

We have analyzed in detail the oscillator algebra of Young diagrams and its
relation to the $u(2)$ algebra by Inonu-Wigner contraction for the large
dimension operators with $p$ long rows or $p$ long columns in the Young
diagrams. The existence of the harmonic oscillator algebra at higher loops
is an important evidence for integrability in this nonplanar regime at
higher loops.

The dependence of the spectra on the $q_{ij}n_{ij}$, where $q_{ij}$ and $%
n_{ij}$ label the oscillator level and the string number respectively, in
the large $N$ limit is protected by the oscillator algebra and is robust
under loop corrections if the oscillator algebra persists at higher loops.

We have provided evidences that the oscillator algebra is preserved in the
large $N$ but nonplanar limit at higher loops. One evidence is that higher
loop dilatation operators are polynomials of the Lie algebra generator which
we call $A^{3},$ in the large $N$ but nonplanar limit, and these operators
will satisfy the oscillator algebra. If in the large $N$ limit, the $h$-loop
dilatation generator preserves the oscillator algebra, then the
integrability in this nonplanar regime is preserved at $h$-loop.

The oscillator algebra itself is a Inonu-Wigner contraction of the $u(2)$
algebra inside of the $u(p)$ algebra of $p$ giant gravitons. Since the $u(p)$
algebra of $p$ number of branes is robust, such symmetry is expected to be
robust under loop corrections.

The existence of BPS states as the lowest energy eigenstates is also one
evidence. Those are the states that can be viewed as the eigenvectors with
zero eigenvalues under the anomalous piece of the dilatation operator. They
are also related to the root of the aforementioned polynomials.

The evidences of the integrability at higher loops in the nonplanar large $N$
limit presented in this paper is in addition to other evidences of different
reasons, which are related to double coset ansatz \cite{deMelloKoch:2012ck}
or to global symmetry \cite{Koch:2013xaa}.

Under higher loop corrections, the spectra are expected to be given by an
interpolating function, that is a function of the 't Hooft coupling as well
as weights of the corner boxes of the Young diagrams. The spring constant of
the oscillator between a pair of giant gravitons will also be an
interpolating function of the coupling constant.

The operators in the oscillator basis can be diagonalized by the harmonic
oscillator wavefunctions and labelled by oscillator levels. Two loop
diagonalizations have been also computed in \cite{deMelloKoch:2012sv}. We
presented evidences that the diagonalization of these operators at higher
loops are not corrected, in the large $N$ limit that we consider, due to the
robustness of both the double coset symmetry and the harmonic oscillator
algebra. We have argued that the higher loop corrections do not change the
basis of the diagonalization of the operators in the large $N$ but nonplanar
limit. They only modify the spectra of these operators by corrections at
higher orders of the coupling.

\vspace{1pt}

\vspace{1pt}

\section*{Acknowledgments}

We would like to thank P. Diaz, T. Harmark, A. Jevicki, Y. Kimura, R. de
Mello Koch, A. Prinsloo, S. Ramgoolam, C. Sieg, J. Simon, R. Suzuki, A.
Torrielli, N. Wallach, S.-T. Yau, and D. Young for interesting relevant
discussions. We would also like to thank Soltis Center, Texas A\&M
University, National Taiwan University, and Queen Mary University of London
for hospitalities. This work was supported in part by NSF grant DMS-1159412,
NSF grant PHY-0937443, NSF grant DMS-0804454, and in part by the Fundamental
Laws Initiative of the Center for the Fundamental Laws of Nature, Harvard
University.





\end{document}